\documentclass[aps,prb,10pt,twocolumn,showpacs,floatfix,superscriptaddress,amsmath,amssymb]{revtex4-1}

\usepackage{inputenc}
\usepackage{url}
\usepackage{graphicx}
\usepackage{dcolumn}
\usepackage{bm}
\usepackage{color}
\usepackage{version}
\usepackage{amsfonts}
\usepackage{soul}
\usepackage{float}
\usepackage{ulem}
\usepackage{hyperref}
\usepackage{mathtools}

\DeclarePairedDelimiter\floor{\lfloor}{\rfloor}

\newcommand{\ba}{\begin{eqnarray}}
\newcommand{\ea}{\end{eqnarray}}
\def\be{\begin{equation}}
\def\ee{\end{equation}}

\excludeversion{details}

\begin{document}


\title{Phase diagram study of a two-dimensional frustrated antiferromagnet via unsupervised machine learning}

\author{S. Acevedo}
\email{santiagoacevedo@fisica.unlp.edu.ar}
\affiliation{IFLP - CONICET, Departamento de F\'isica, Universidad Nacional de La Plata, C.C.\ 67, 1900 La Plata, Argentina.}

\author{M. Arlego}
\affiliation{IFLP - CONICET, Departamento de F\'isica, Universidad Nacional de La Plata, C.C.\ 67, 1900 La Plata, Argentina.}

\author{ C. A.\ Lamas}
\affiliation{IFLP - CONICET, Departamento de F\'isica, Universidad Nacional de La Plata, C.C.\ 67, 1900 La Plata, Argentina.}

\begin{abstract}

We apply unsupervised learning techniques to classify the different phases of the $J_1-J_2$ antiferromagnetic Ising model on the honeycomb lattice.
We construct the phase diagram of the system using convolutional autoencoders.
These neural networks can detect phase transitions in the system via `anomaly detection', without the need for any label or a priori knowledge of the phases.
We present different ways of training these autoencoders and we evaluate them to discriminate between distinct magnetic phases.
In this process, we highlight the case of high temperature or even random training data.
Finally, we analyze the capability of the autoencoder to detect the ground state degeneracy through the reconstruction error.

\end{abstract}

\maketitle

\section{Introduction}

Machine learning techniques have been applied to condensed matter systems in a wide variety of contexts and tasks, as
detecting phase transitions from synthetic \cite{carrasquilla2017machine,alet2019,MBL2018,MLdynamics2018,DasSarma-topo-2017,Zhang-topo2018,Melko2017-interpretable-kernel,corte2020transfer}
and experimental data\cite{expML2019,expML2020},
improving the efficiency of Monte Carlo sampling in classical and quantum systems\cite{liu2017montecarlo,xu2017Qmontecarlo,Wangmontecarlo2017},
quantum state tomography\cite{torlai2018tomography}, determination of critical exponents\cite{crit2019,GIANNETTI2019crit},
modeling thermodynamic observables for physical systems in thermal equilibrium\cite{Melko-RMB-thermo2016}, encoding many-body quantum
states\cite{carleo2017wavefunction}, etc.
As a result, machine learning techniques have gained a significant place in the subject of interacting many-body condensed matter physics.

In several disciplines, the classification problem plays a central role.
The classification of the types of objects in an image or text and the different phases in a condensed matter system are tasks that share common challenges.
However, the tools in both cases are different in nature.

In condensed matter, the classification of phases historically relies on the use of order parameters, which allow differentiating the different phases
present in a model. Nonetheless, this approach demands detailed knowledge of the system, which may not be available, particularly in many-body quantum
systems of current interest, such as spin liquids or materials with topological properties.

In the machine learning community, a whole battery of techniques specifically oriented to classification based on data has been developed in recent years.
There are several types of classifiers in machine and deep learning. Broadly speaking, they can be differentiated by the degree of a priori knowledge required. At one extreme, the supervised algorithms, require prior labeling on a set.
This is used to train a neural network (which can be of different types, dense, convolutional, recurrent, etc), that learns to classify a data set.
An interesting property here is the generalization scope of the neural networks, which in certain circumstances can make correct predictions 
beyond the original sets of data with which they were trained.
Other methods, called semi-supervised, require partial knowledge about the data, without complete labeling\cite{van2020survey}.
%
Finally, at the other extreme are the non-supervised methods, which require no prior knowledge of the system and build knowledge directly from the data.
Currently, there are a variety of unsupervised methods. These can be classified according to the type of architecture used, i.e. shallow or
deep, linear or non-linear, etc. On the one hand, Principal Component Analysis (PCA)\cite{jolliffe1986principal} and  Uniform Manifold Approximation
and Projection (UMAP)\cite{mcinnes2020umap}, are examples of shallow, linear and non-linear dimensionality reduction techniques, respectively. Furthermore, diffusion maps\cite{Coifman-diff} constitute a shallow non-linear algorithm well-suited to study topological\cite{rodriguez2019topological} and quantum\cite{Diffusion-maps2} phase transitions in many body systems.
On the other hand,  convolutional and variational autoencoders\cite{DeepLearningBible} provide powerful tools for non-linear unsupervised learning, based on deep neural networks architectures.

In this work we determine the phase diagram of a paradigmatic case of frustrated magnetism in two dimensions,
the Ising model on the honeycomb lattice employing unsupervised machine learning.
The studies of the two-dimensional Ising model has a long story on the square lattice.
In this geometry, the interplay between frustration and temperature gives rise to phase transitions whose nature has 
been the subject of debate \cite{grynberg1992square,kalz2009monte,kalz2008phase,Andreas-Frustrated-Ising-Square-1}.
Comparatively, the phase diagram of the frustrated honeycomb lattice has been less explored.
This lattice is interesting both from a theoretical and an experimental point of view.
From the theoretical point of view, the honeycomb lattice has the smallest coordination number for a two-dimensional lattice ($z=3$).
For this reason, it is expected that the correlations are strong in this system.
The small coordination number together with frustrated interactions has been shown to have interesting effects in the Heisenberg model on this lattice,
where the phase diagram shows a magnetically disordered region and a spin-liquid phase\cite{PhysRevB.87.024415,PhysRevB.83.094506}.
From the experimental point of view, several materials regarded as realizations of spin systems on honeycomb lattices have
being synthesized \cite{KATAEV2005310,smirnova2009synthesis,tsirlin2010beta,matsuda2010disordered,okubo2010high}.
Therefore, the complexity and interest of this model make it a candidate for testing the ability of unsupervised techniques to detect transitions in
two-dimensional frustrated systems.

The plan of the manuscript is as follows.
For completeness, in Sec. \ref{sec:ising} we present a brief discussion of the prominent aspects of the Ising model in the presence of frustration,
and in Sec. \ref{sec:intro_unsupervised} we introduce the main features of autoencoders.
%
%
In Secs. \ref{sec:tow_T}, and \ref{sec:high_T} we train autoencoders to discriminate the distinct phases of the system.
In Sec. \ref{sec:PD} we construct the entire phase diagram of the system in an unsupervised way.
Furthermore, here we show how to use autoencoders to generate labels that can be used to perform classifications with supervised learning techniques.
In Sec. \ref{sec:conclusions} we present a summary of results and conclusions.

\begin{figure}[t!]
\begin{center}
  \includegraphics[width= .7\linewidth]{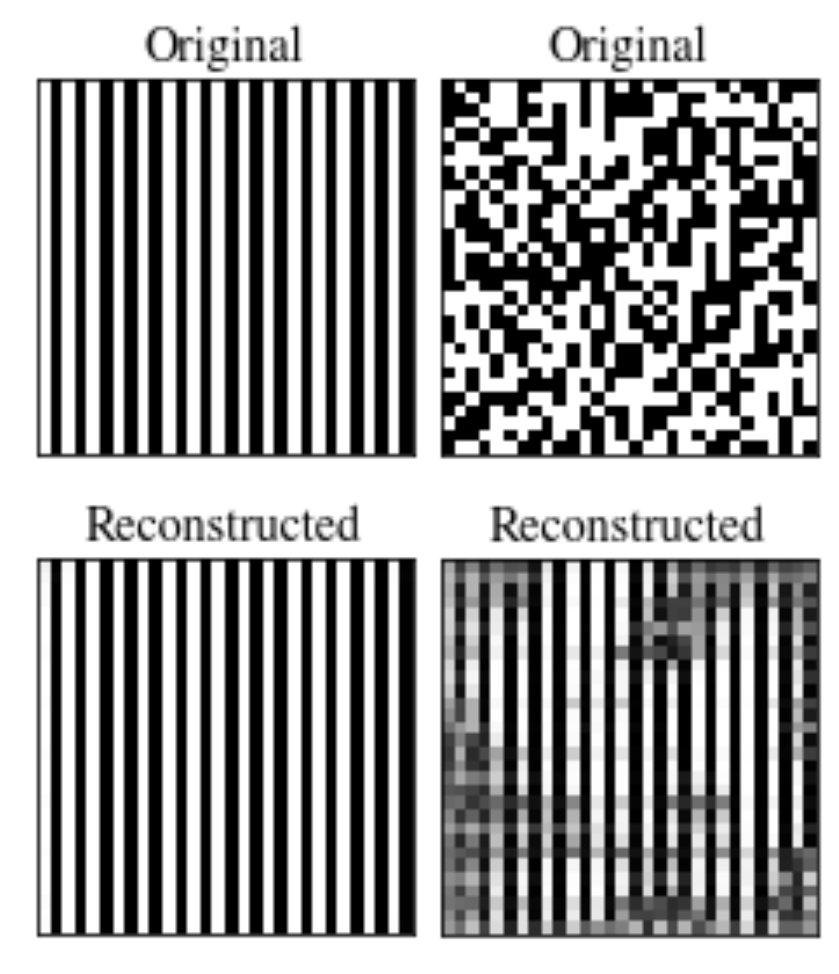}
 \caption{Upper row: Original snapshots from the system mapped to an square array as detailed in appendix \ref{sec:honey_map}. White and black pixels correspond to up and down spins, respectively.
  On the left, a N\'eel ordered snapshot corresponding to $J_2=0$ and $T=0.02$. On the right, a paramagnetic snapshot corresponding to $J_2=0$ and $T=4.5$. Lower row: Respective reconstructions using the CAE from training I.}
 \label{fig:reconstructions_Neel}
\end{center}
\end{figure}

\subsection{Model}
\label{sec:ising}

We study the $J_1-J_2$ antiferromagnetic Ising model on the honeycomb lattice, in the absence of an external magnetic field.
Each spin is represented with a discrete variable $\sigma_{i}=\pm 1$, defined on the lattice. It interacts with its first and second neighbors with
antiferromagnetic couplings $J_1$ and $J_2$, respectively. The system Hamiltonian is then

\begin{equation}
  \label{eq:Ising-Hamiltonian}
 H=J_1 \sum_{ \langle i,j\rangle } \sigma_i\sigma_{j} + J_2 \sum_{ \langle \langle i,j\rangle \rangle } \sigma_i\sigma_{j}
\end{equation}
with $J_1,J_2>0$. In this work we take $J_1=1$, fixing the energy scale. Then $J_2>0$ regulates the
frustration in the system. Furthermore, the Boltzmann constant is absorbed in $T$, and then $T$ is also measured in units of $J_1$.  \\
Frustration in many-body magnetic systems corresponds to the impossibility to simultaneously minimize all interactions, leading to high degeneracy.
This can produce a variety of behaviors, even in simple models as the Ising one. As a first example, we can consider the antiferromagnetic Ising model
on the square lattice. For 
$J_2 > 1/2$ the ground state is four-fold degenerate and presents a stripe-type long-range order.
Furthermore, the nature of the phase transition between the ordered and disordered phases is a long-term
debate \cite{Andreas-Frustrated-Ising-Square-1,Andreas-Frustrated-Ising-Square-2,Ising-square-Coll-Sandvik-1,Ising-square-Coll-Sandvik-2}.

\begin{figure}[t!]
\begin{center}
  \includegraphics[width= \linewidth]{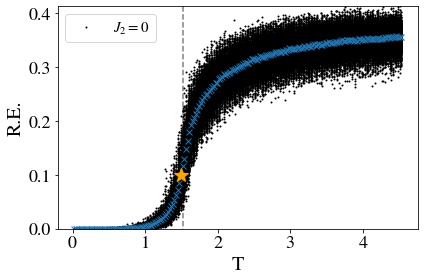}
 \caption{Reconstruction Error (R.E.) vs temperature for 400 independent realizations of the unfrustrated system ($J_2=0$), computed with the CAE from training I. The dashed gray vertical line corresponds to the analytical critical temperature of the system in the thermodynamic limit and the yellow star corresponds to the estimation of the inflection point in the mean R.E. curve, denoted by the blue crosses.}
 \label{fig:errors_J2_0}
\end{center}
\end{figure}

As a second example, we mention the antiferromagnetic nearest neighbors Ising model on the triangular lattice.
The geometrical frustration in this lattice introduces a macroscopic degeneracy which destroys long-range order even at $T=0$.
The system presents zero-point entropy and has no Curie point\cite{wannier1950-Ising-AFM-Triangular}.
Nonetheless, even being a paramagnet at all temperatures, 
some correlations emerge at low temperatures.
In this way, the system can be seen as a classical spin liquid\cite{balents2010spin}.

The honeycomb lattice, dual to the triangular lattice, is not geometrically frustrated.
Here two different sub-lattices can be defined, in which a conventional antiferromagnetic or N\'eel order fits. For $J_2=0$,
both the ferromagnetic and the antiferromagnetic Ising models have a critical temperature in the thermodynamic limit that can be determined
exactly\cite{Ising-Exact-Honeycomb} and is given by $T_c=2 /\log{(2 + \sqrt{3})} \approx 1.519 $.

At $J_2=1/4$ this system presents a low-temperature phase transition from a N\'eel phase to a degenerate phase without
long-range order\cite{Ising-Honeycomb-frustrated}.

In appendix \ref{sec:PCA} we present an introduction to principal component analysis and its aplication to our model, where some of the central features of the system can be detected by its linear low dimensional representations, like
i) the presence of a low-temperature phase with a single linear order parameter, and a 2-degenerated ground state, i.e., the N\'eel phase;
ii) the absence of a phase with previous characteristics for $J_2>1/4$,
iii) the presence of a high-temperature uncorrelated phase, and
iv) signals of the value $J_2=1/4$ as the maximally frustrated point of the system. Nonetheless, to construct the phase diagram of the system through unsupervised machine learning  is necessary to utilize more powerful non linear techniques, as autoencoders.

\subsection{Autoencoders}
\label{sec:intro_unsupervised}

 \begin{figure}[t!]
 \begin{center}
 \includegraphics[width= .8\linewidth]{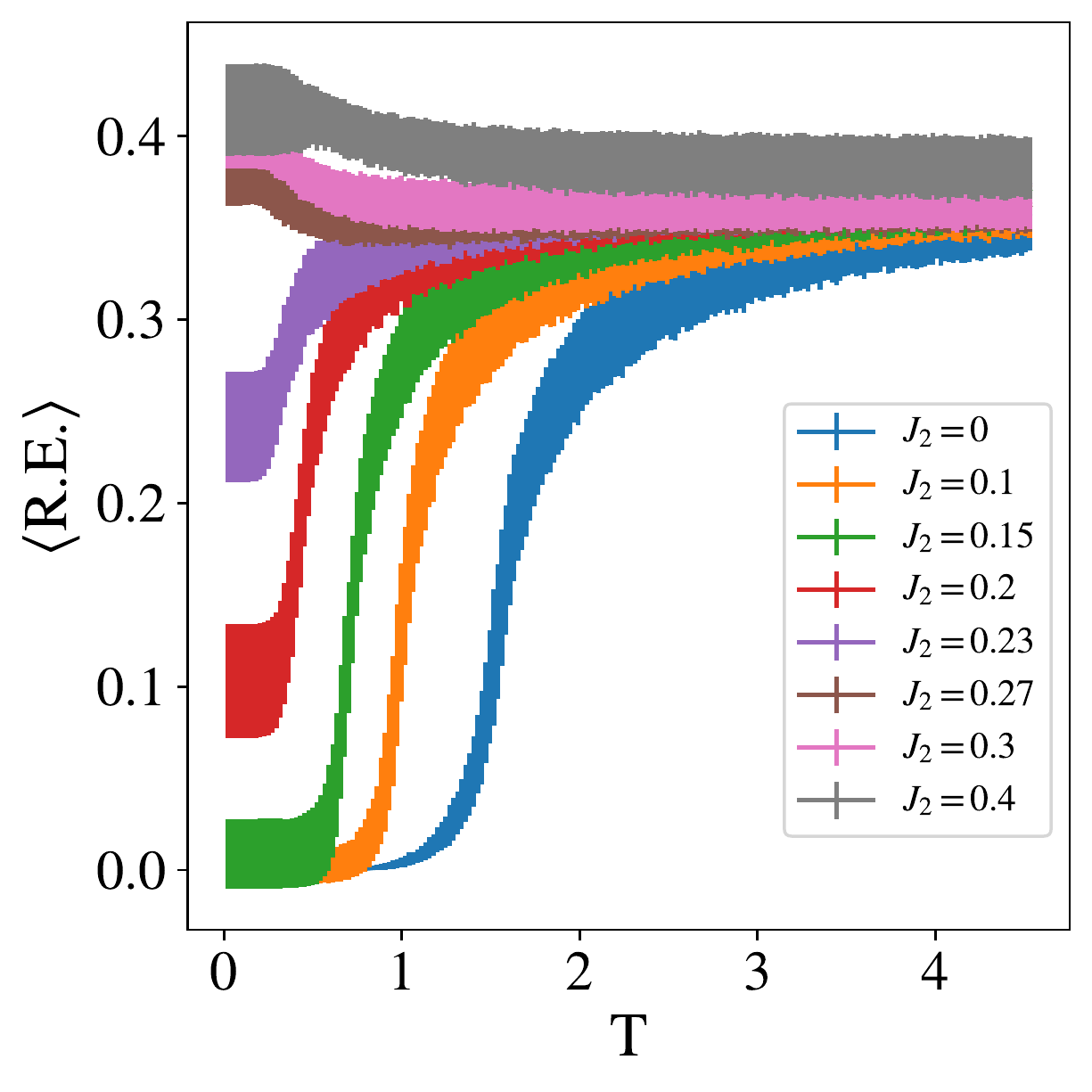}
 \caption{Phase discovery: Mean reconstruction errors vs temperature for several values of $J_2$, computed with the CAE corresponding to training I. Error bars show the standard deviation computed with the 400 realizations of the system.}
 \label{fig:RE_neel}
\end{center}
\end{figure}



Autoencoders (AEs) are a particular type of neural network that is trained to generate an approximate copy of the given input, $x$.
They are composed of two parts\cite{DeepLearningBible}: the encoder, which codifies $x$ into a code $h=f(x)$ that belongs to the so-called latent space, and the decoder, which takes the code and tries
to reproduce the original input via a mapping $g(h)=\hat{x}\approx x$.
The AE trainable parameters are adjusted minimizing a loss function that measures the difference between $x$ and its reconstruction $\hat{x}$, for each x in a given training set.

In this work we choose the standard loss function that measures the distance between $x$ and $\hat{x}$, the mean square error $(mse)$. For the case of a set $X$ of $N$ two-dimensional input images $x^n$ of size $L\times L$, the reconstruction error of this set is given by 
\begin{equation}
 mse(X)= \frac{1}{N L^2} \sum_{n=1}^N\sum_{i=1}^L\sum_{j=1}^L |x^n_{ij}-\hat{x}^n_{ij}|^2.
\end{equation}
In the case of systems with a simple order parameter, as non-frustrated Ising models, it can be shown that PCA or
autoencoders with a single neuron in the latent space  encode the spin configurations in a variable $Z$ that is highly correlated with the order
parameter of the system\cite{wetzel2017unsupervised}.
Nonetheless,  at high temperatures or doing finite-size scaling these two quantities differ\cite{alexandrou2019}.

The layers in the AE can be dense (fully-connected neurons) or convolutional.
In this work, we focus on AEs that are completely convolutional, i.e., convolutional autoencoders (CAEs). The details of the architectures used are presented in appendix \ref{sec:arquitecturas}.
They have the advantages of having exponentially fewer parameters than dense or fully-connected AEs, and the ability to capture spatial correlations
thanks to the application of filters, or local transformations, throughout the entire image.
CAEs have been successfully applied in a many-body quantum system\cite{kottman2020} to construct the phase diagram of the Bose-Hubbard model in
an unsupervised way. It is important to remark that these CAEs do not have a latent space of a few variables as in Refs. \onlinecite{wetzel2017unsupervised} and \onlinecite{alexandrou2019}, and therefore our objective implementing them is not to generate
low dimensional representations of data, but to predict phase transitions via the concept of anomaly detection.
This concept consists of training an AE to reproduce a certain class of data until
the reconstruction error (R.E.) is small enough.
Then, if a different class of data is given to the AE as input, usually the reconstruction fails, because the AE was not trained to reproduce it,
and the R.E. augments abruptly, signaling the anomaly.


\begin{center}
 \begin{figure}[t!]
 \includegraphics[width= \linewidth]{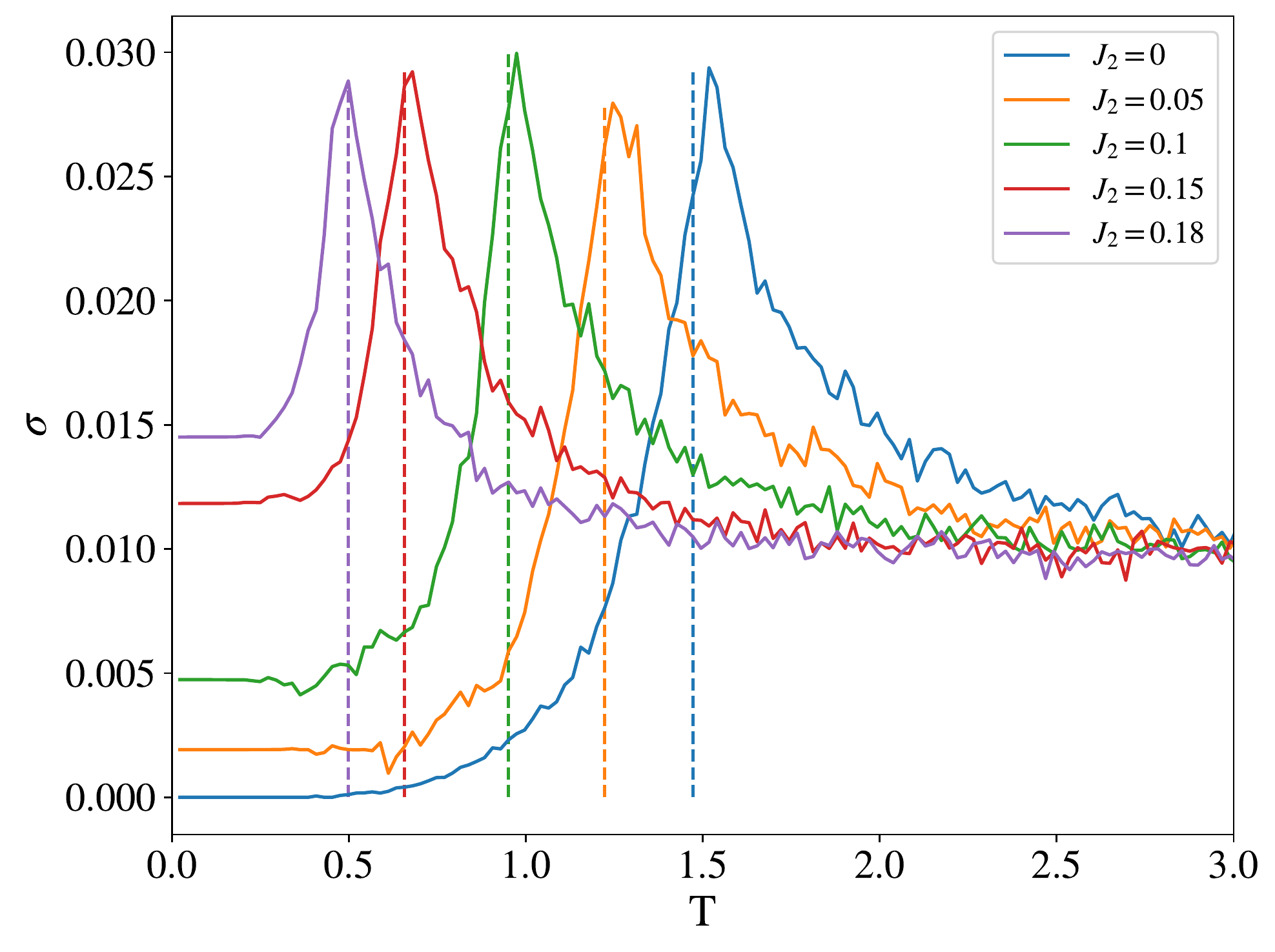}
 \caption{Standard deviation $ \sigma$ in the R.E. as a function of temperature, for different values of $J_2$ in the N\'eel phase, computed with the CAE corresponding to training I. In dashed lines, the estimation of the inflection points in each mean R.E. curve.
 }
 \label{fig:dispersiones_neel}
\end{figure}
\end{center}

\begin{figure}[t!]
\begin{center}
  \includegraphics[width= \linewidth]{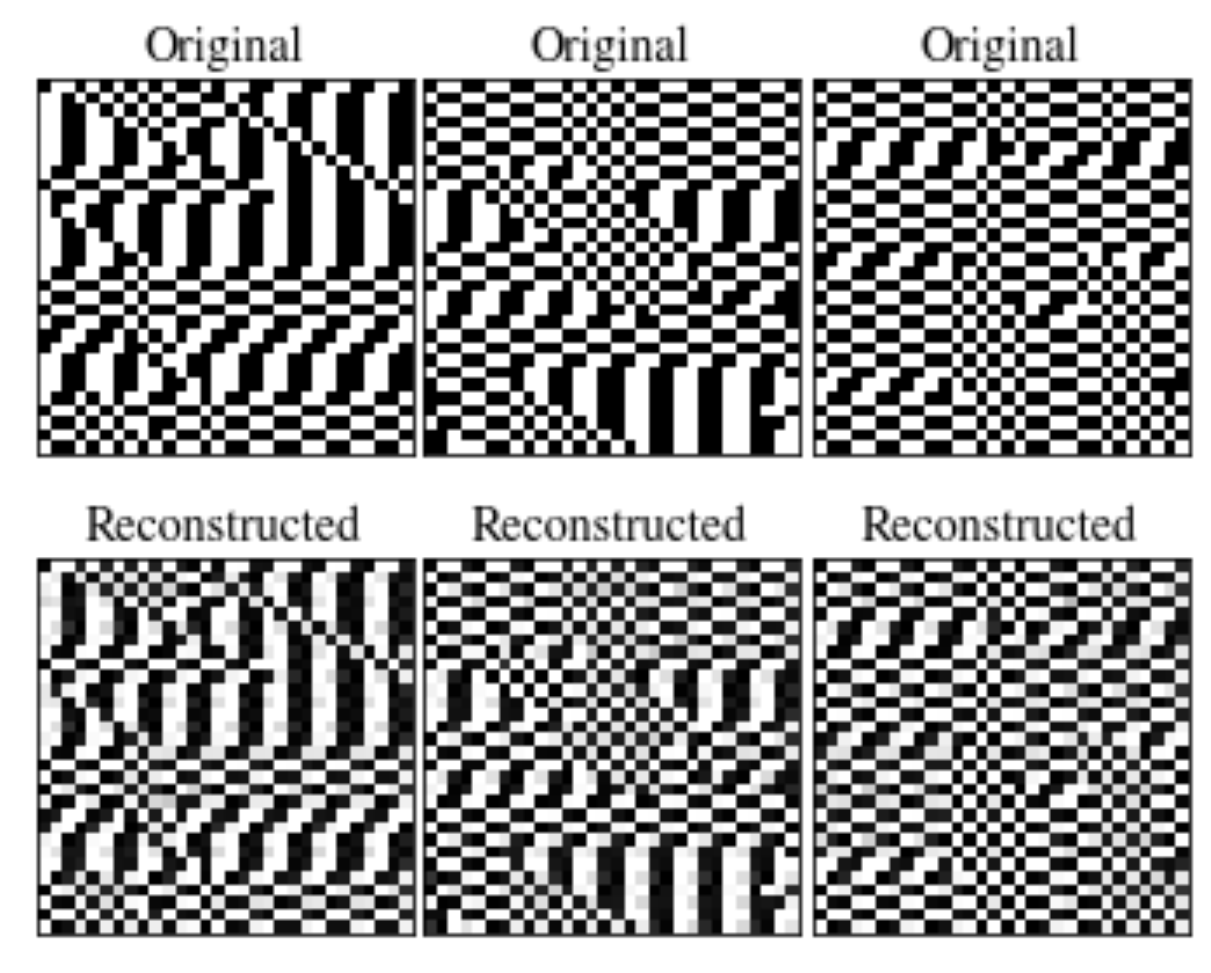}
 \caption{Upper column: three snapshots corresponding to temperatures $T=0.02$ for three realizations of the system with $J_2=1/2$. Lower column: reconstructions made by the CAE corresponding to training II.}
 \label{fig:frustrated_snapshots}
\end{center}
\end{figure}

\section{Results}


To test the predictive power of the CAEs, we make different choices of training sets of data and analyze the respective performances.
The system size in this work is fixed to $N=900$ sites, and the Monte Carlo information is presented in appendix \ref{sec:MC}.
The details of the autoencoder architecture and hyperparameters are presented in appendix \ref{sec:arquitecturas}.
Below we detail the results obtained with each training.

\subsection{Learning at low temperatures}
\label{sec:tow_T}

\begin{figure}[t!]
 \includegraphics[width= \linewidth]{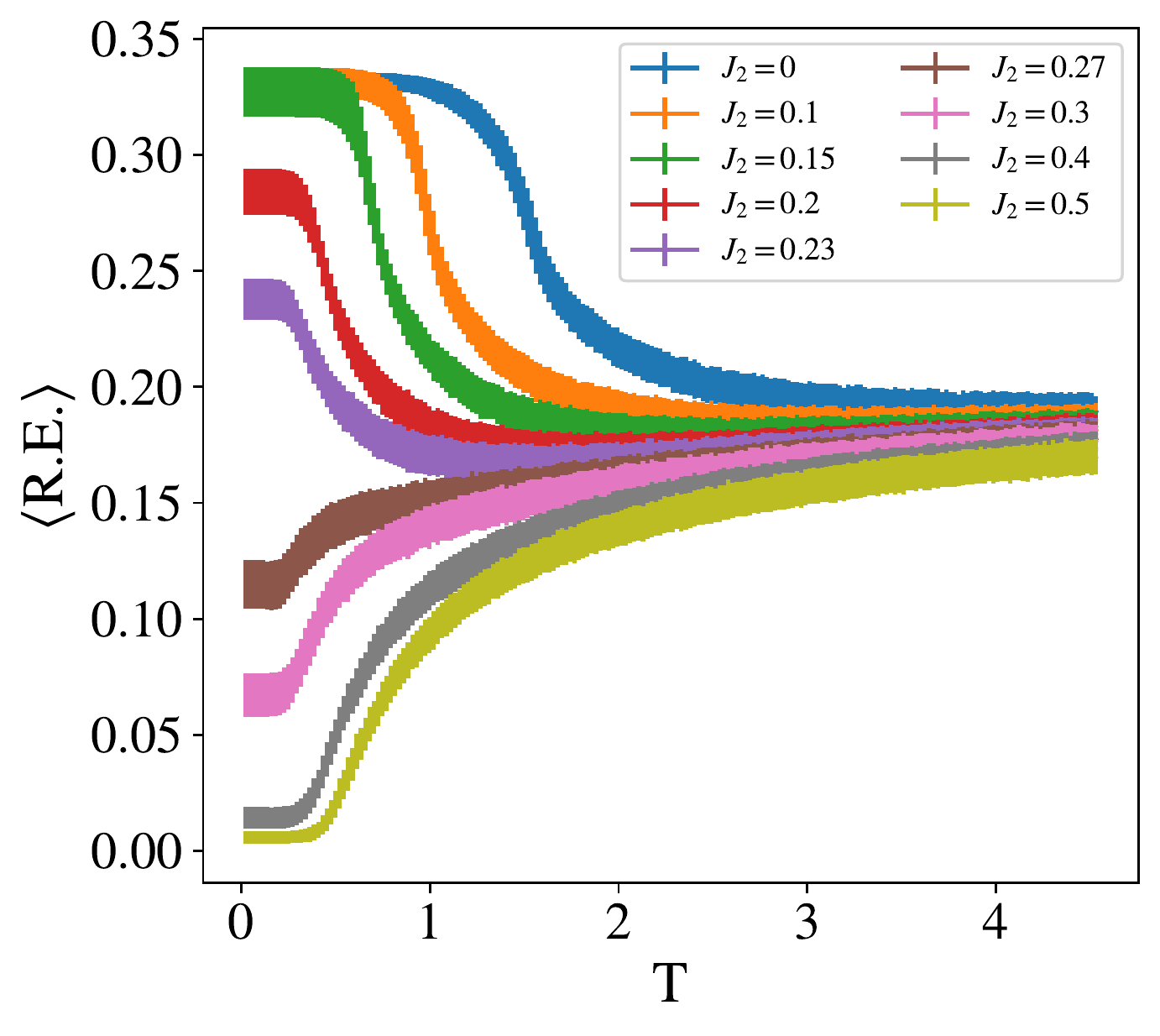}
 \caption{Phase discovery: Mean reconstruction errors vs temperature for several values of $J_2$, computed with the CAE corresponding to training II. Error bars show the standard deviation computed with the 400 realizations of the system.}
 \label{fig:RE_ice}
\end{figure}

\subsubsection{Training I: unfrustrated low-temperature training data} 

We trained a CAE to reconstruct N\'eel ordered spin configurations corresponding to $0.02<T<0.2$ and $J_2=0$. We denote this 
as training I.  These configurations, that constitute the CAE's input, can be in a variety of formats. In this work, we map the honeycomb lattice to a square array as detailed in appendix \ref{sec:honey_map}.
Fig. \ref{fig:reconstructions_Neel} upper left and right panels show snapshots of the system with only nearest neighbors interactions $(J_2=0)$, at $T=0.02$ and $T=4.53$, respectively. Note that the N\'eel order in the honeycomb lattice is mapped to a stripe order in the square array.
Fig.  \ref{fig:reconstructions_Neel} lower panels show the corresponding reconstructions.
The CAE can reproduce the N\'eel ordered spin configuration (R.E. of order $10^{-4}$) and tries to generate a N\'eel ordered spin configuration even from the disordered snapshot.
The maximum possible R.E. within our choices of normalization and metric is unity, as explained in appendix \ref{sec:arquitecturas}.

Fig. \ref{fig:errors_J2_0} shows the R.E. as a function of temperature for the 400 realizations of the system.
As the system converges to one of the two possible N\'eel ground states at low temperatures, the CAE is able to reproduce the corresponding
spin configuration and the R.E. reaches the minimum value (order $10^{-4}$).  At high temperatures, the CAE does not reproduce the disordered spin
configurations, as shown in Fig. \ref{fig:reconstructions_Neel}, and the R.E. increases. As it can be observed, there is a very good agreement between the mean R.E inflection point and the critical temperature of the system in the thermodynamic limit. The numerical estimation of the R.E. inflection point is explained in appendix \ref{sec:SG}.

Fig. \ref{fig:RE_neel} shows the mean R.E. as a function of temperature for different values of $J_2$. At high temperatures, all curves tend to converge
to a constant mean value, as the system is in the paramagnetic phase. At low temperatures, R.E. decreases with respect to the high-temperature value for
systems with $J_2<1/4$, which are in the N\'eel phase. This is not the case for systems with $J_2>1/4$, in agreement with the low-temperature phase transition at $J_2=1/4$.

Albeit in Fig. \ref{fig:RE_neel} it is possible to distinguish three regimes that correspond to the three phases of the system, 
for the curves with $J_2>1/4$ the R.E. does not allow to discriminate properly the high and low temperature regimes. 
For example, for the gray curve ($J_2=0.4$) in Fig.\ref{fig:RE_neel}  the low and high temperature R.E.s coincide within dispersion.
This is can be understood as a consequence of the training simplicity.
The CAE had the task to reproduce N\'eel  ordered configurations, and with that knowledge, it is being used to try to distinguish highly frustrated configurations from disordered configurations, both of which do not present long range order. Nonetheless, this difficulty is overcome in training II.

In Fig. \ref{fig:errors_J2_0} it is clear that the dispersion is also a function of temperature, due to the presence of an ordered phase and a 
disordered phase.
Fig. \ref{fig:dispersiones_neel} shows the standard deviation $\sigma$ in the R.E. as a function of temperature, for different $J_2$ values. 
For $J_2<1/4$ there are peaks in the dispersion that coincide with the respective inflection points, denoted by the vertical dashed lines. 
Additionally, it can be seen that all dispersions converge approximately to the same value at high temperatures. At low temperature, in contrast, 
the dispersion augments with increasing frustration.

For $J_2>1/4$ it can be seen that the dispersion augments concerning the high-temperature value around the transition temperatures,
but there are no peaks as in Fig. \ref{fig:dispersiones_neel}.
A figure that shows this behavior will be presented in Sec. \ref{sec:trainingIV}, for a different training choice. 

\begin{figure}[t!]
 \includegraphics[width= \linewidth]{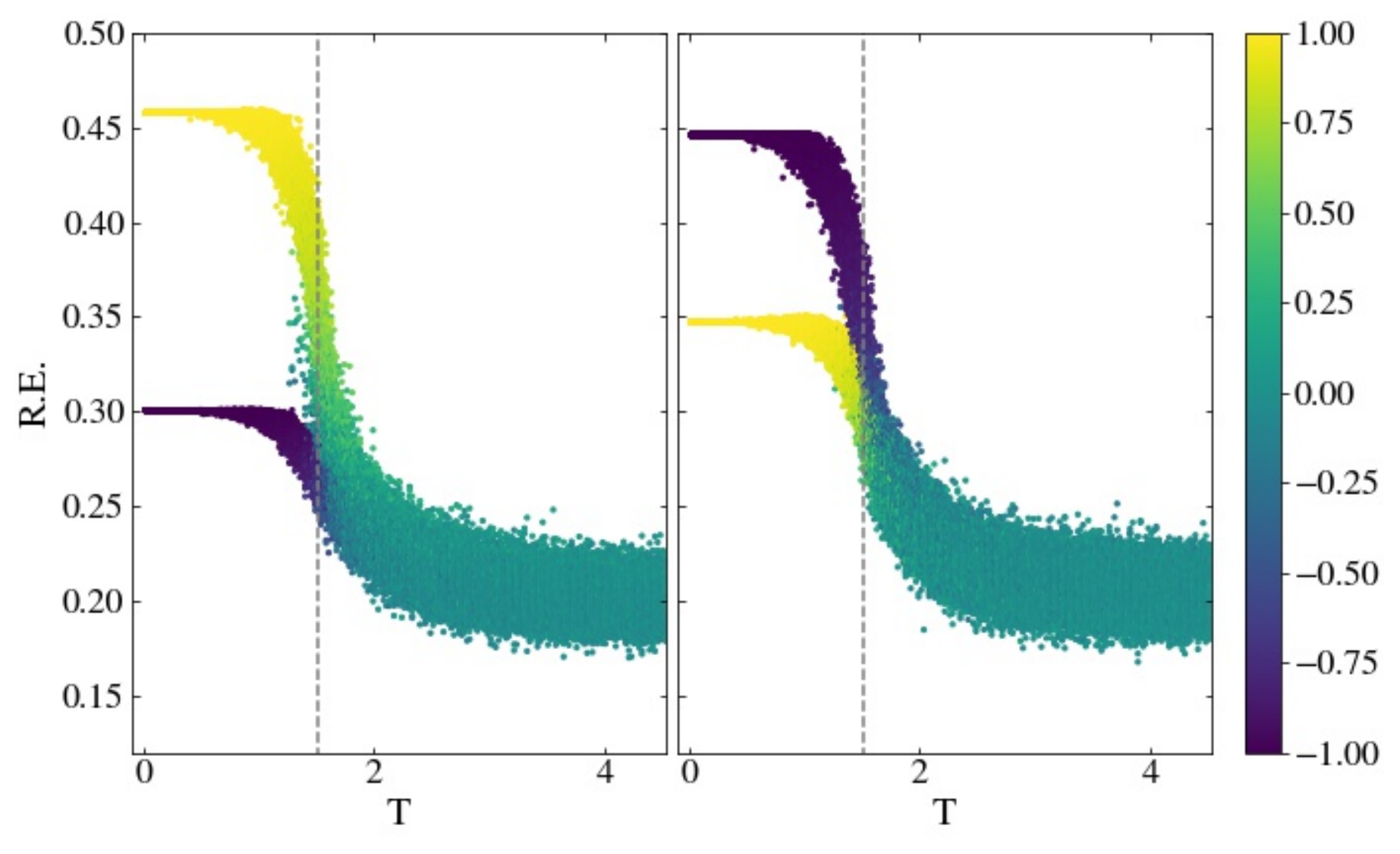}
 \caption{Reconstruction error vs temperature for the 400 realizations of the unfrustrated system ($J_2=0$), computed with the CAE corresponding to training II. The color scale shows the staggered magnetization per site $m$ of each snapshot, i.e., the order parameter in the N\'eel phase. Each panel corresponds to an equivalent but independent training, showing that each N\'eel ground state can have the lowest R.E. at the splitting.
}
 \label{fig:colored_spliting}
\end{figure}

\subsubsection{Training II: frustrated low-temperature training data}

We trained a CAE with spin configurations with $0.02<T<0.2$ and $J_2=1/2$, where the system has no long range order and a highly degenerate ground 
state due to frustration\cite{Ising-Honeycomb-frustrated}. We denote this 
as training II.
Fig. \ref{fig:frustrated_snapshots} shows  snapshots  at $T=0.02$ for three realizations of systems with $J_2=1/2$ in the upper row, and the 
corresponding reconstructions in the lower row, where it can be seen that there are zones of different local orders. The final R.E. in validation data
during training is order $10^{-3}$.

Fig. \ref{fig:RE_ice} shows the mean R.E. as a function of temperature for several values of $J_2$. The behavior is similar to that 
of Fig. \ref{fig:RE_neel}. As the system remains in the phase in which the autoencoder was trained, the R.E. decreases at low temperatures, relative to 
the shared high-temperature value.

Fig. \ref{fig:colored_spliting} shows the R.E. computed for the 400 realizations of systems with $J_2=0$. The color scale corresponds to the
staggered magnetization per site $m$, i.e., the order parameter in the N\'eel phase. 
Each panel corresponds to a different but equivalent training, and the vertical dashed line corresponds to the analytical critical temperature of the system.
In this case the R.E. not only increases, signaling the phase transition, but also it splits in two parts. The additional information of the order
parameter allows to prove that the splitting corresponds to the $Z_2$ symmetry of the Hamiltonian, broken in the ground state. Nonetheless, 
there should be no a priori preference for neither of the two N\'eel configurations to have  the lowest R.E.  For this reason we show that independent
trainings can give the two possible outcomes.

\begin{figure}[t!]
 \includegraphics[width=250pt]{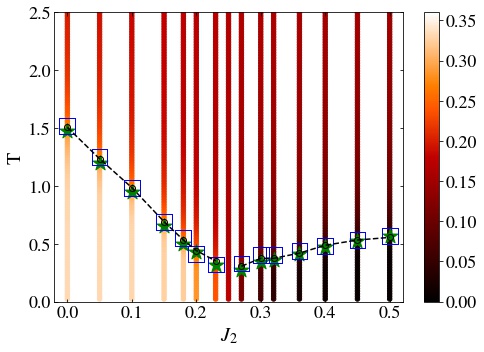}
 \caption{
 Phase Diagram of the system. In color scale the R.E. computed with the CAE corresponding to training II. 
 In green stars, the inflection points of each mean R.E. curve in Fig. \ref{fig:RE_ice}. In black circles joined with dashed lines the transition 
 temperatures obtained with CNNs that used the green stars as labels (see text). In blue boxes the specific heat maxima  in the Monte Carlo simulations.
 }
 \label{fig:PD}
\end{figure}

\subsection{Phase Diagram}
\label{sec:PD}

Fig. \ref{fig:PD} shows the $T-J_2$ phase diagram of the system, constructed by the CAE corresponding to training II, where the color scale denotes 
the R.E. The upper part of the diagram corresponds to the paramagnetic phase, where the R.E. is approximately the same for all values of $J_2$, as shown in
Fig. \ref{fig:RE_ice}.
The bottom right (left) corner of the diagram corresponds to the curves in Fig. \ref{fig:RE_ice} with $J_2>1/4$ ($J_2<1/4$), that have a decrease (increase)
in their R.E. with respect to the high-temperature value.
Green stars correspond to the mean R.E. inflection point for each value of $J_2$ in Fig. \ref{fig:RE_ice}, which are taken as the boundaries between
different phases, generalizing the results from Fig. \ref{fig:errors_J2_0}.

To check if the inflection points estimate correctly the location of the phase transitions, we perform classifications
using a convolutional neural network (CNN) in which we label the data using this quantity as follows:
Given a value of  $J_2$, we take spin configurations that are far away from the temperature corresponding to the inflection point, $T^*$, i.e.,
that have $|T-T^*|>w=0.2$. We label the high-temperature configurations with 1 and the low-temperature configuration with 0 and we train a CNN with these
configurations.  As the resulting validation accuracy is higher than 0.999, we can ensure that the transition is within the temperature window extracted,
i.e., near $T^*$. It is known that wrong labeling would lead to a diminishing in the classification accuracy\cite{confusion}.

\begin{figure}[t!]
\begin{center}
  \includegraphics[width=  \linewidth]{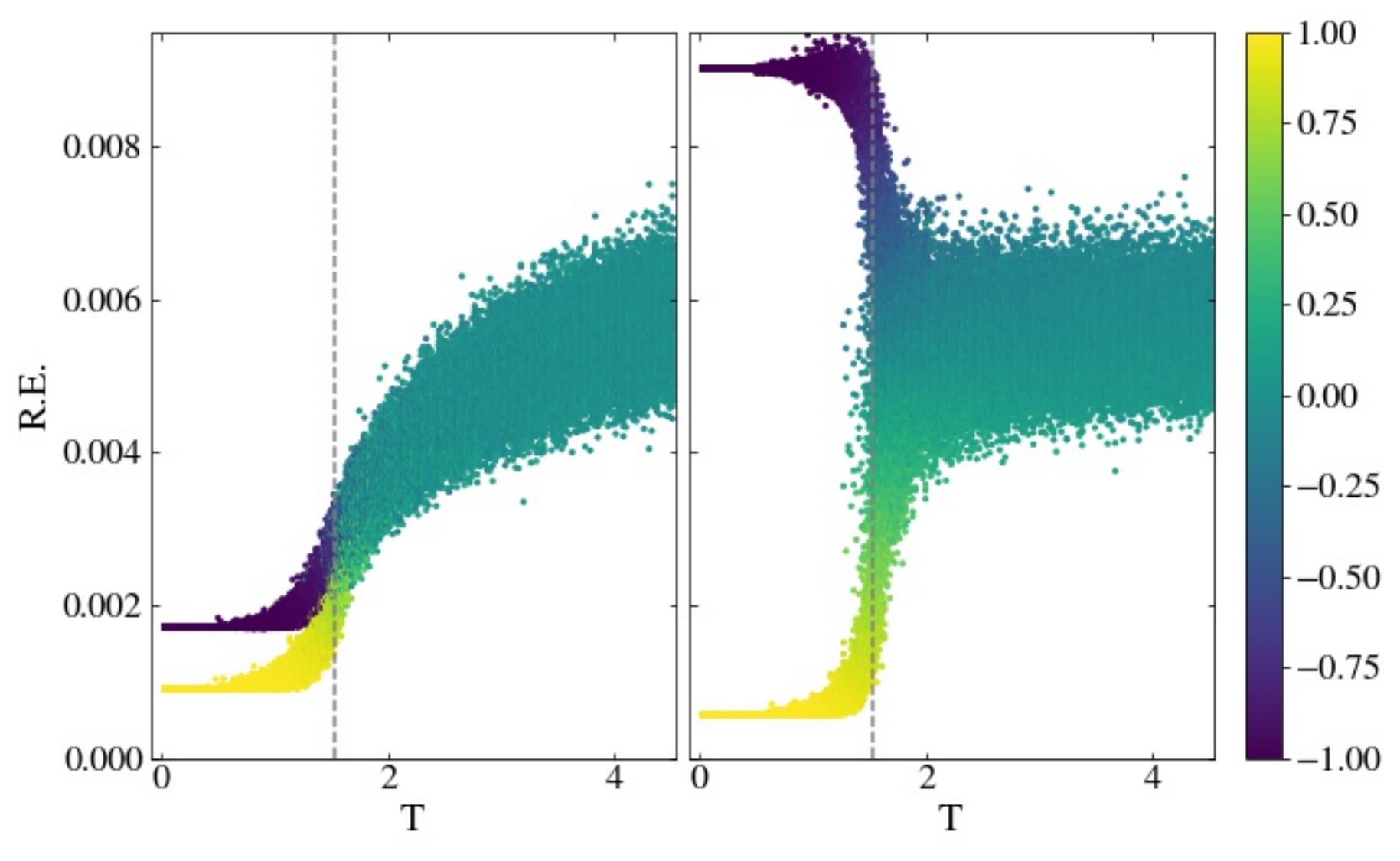}
 \caption{
 Reconstruction error vs temperature for the 400 realizations of the system with $J_2=0$. The left (right) panel was computed with the CAE corresponding
 to training III (IV).
 The vertical dashed line corresponds to the analytical critical temperature of the system. The color scale corresponds to the staggered magnetization 
 per site $m$, i.e., the order parameter in the N\'eel phase.
}
 \label{fig:noisy}
\end{center}
\end{figure}

Then, we make the CNN predict the order-disorder probabilities for spin configurations in the whole range of temperatures, and we find the temperature
at which the two probabilities cross each other. The corresponding transition temperatures obtained are plotted in Fig. \ref{fig:PD} in black empty circles
joined by dashed lines.
Albeit in these classifications we are using supervised learning architectures, the labeling is obtained using the previous unsupervised results from
Fig. \ref{fig:RE_ice}. For this reason, this second prediction of the transition temperatures can be still considered as unsupervised learning, similarly
to the confusion method\cite{confusion}.
Finally, for comparison, blue empty squares in Fig. \ref{fig:PD} correspond to the maximum value of the specific heat of the system,
which signals the transition for each $J_2$. The CNN architecture and hyperparameters are presented in appendix \ref{sec:arquitecturas}.

\subsection{Learning at high temperatures}
\label{sec:high_T}

\subsubsection{Training III: high-temperature training data} 

Here we train a CAE with spin configurations with $J_2=0$ and temperatures  $4<T<4.53$, corresponding to the paramagnetic phase.
We denote this training as training III. \\
The left panel of Fig. \ref{fig:noisy}  shows the R.E. as a function of temperature for spin configurations with $J_2=0$.
Although the CAE was trained in the paramagnetic phase, it can learn the low-temperature order and it even shows the splitting as in Fig. \ref{fig:colored_spliting}. 
It is important to remark that the reciprocal situation does not occur neither in training I nor in training II, i.e., when a CAE learns a low-temperature phase it is unable to reconstruct  disordered snapshots, which is a a harder task.

\begin{figure}[t!]
\begin{center}
  \includegraphics[width= \linewidth]{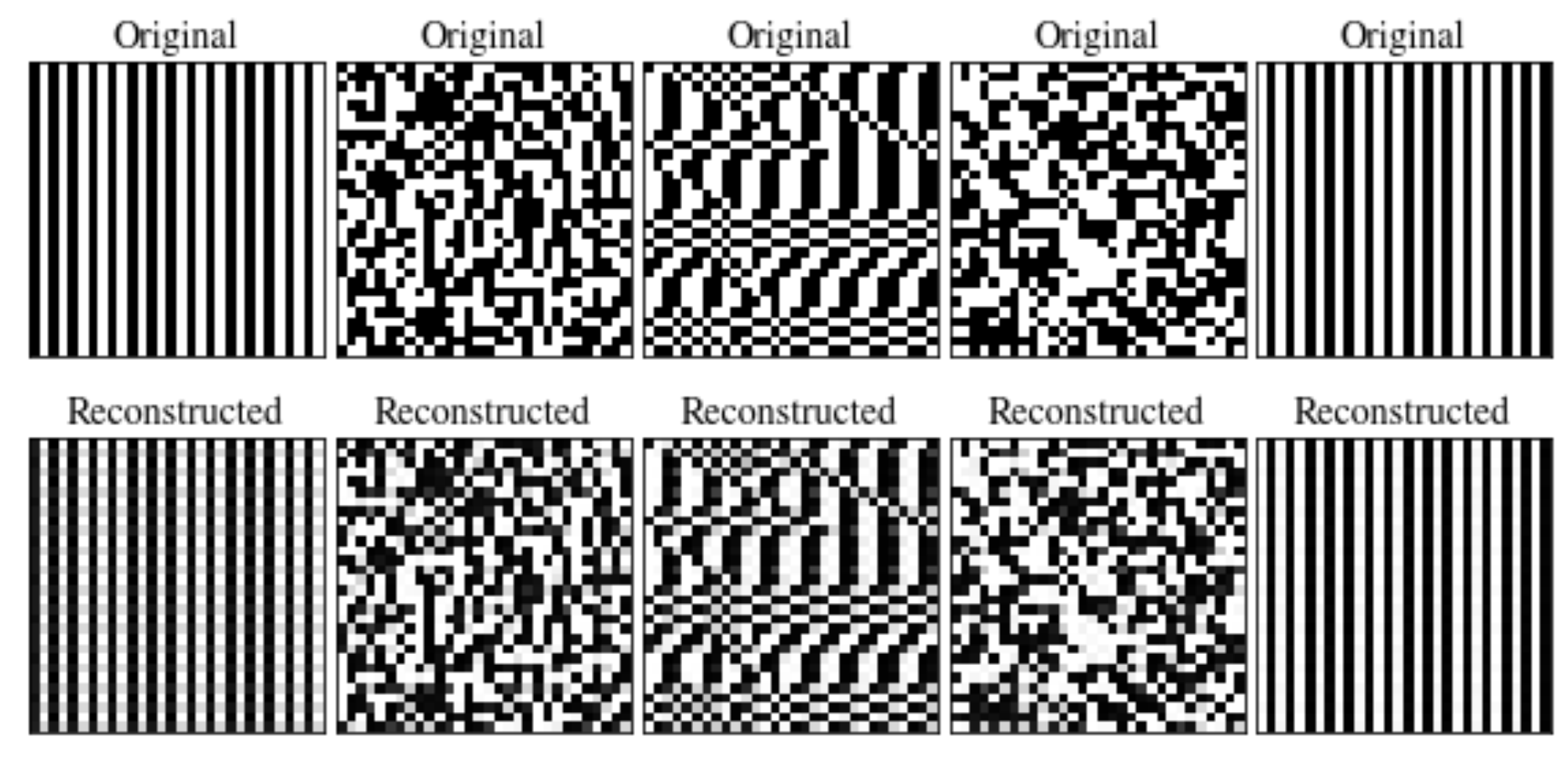}
 \caption{
Reconstructions computed with the CAE corresponding to training IV. From left to right, a N\'eel ordered snapshot with $m=-1$, $J_2=0$ and $T=0.02$, 
 a paramagnetic snapshot with $J_2=0$ and $T=4.53$, a low-temperature frustrated snapshot with $J_2=1/2$ and $T=0.02$, a synthetic random (infinite temperature) snapshot,  and a N\'eel ordered snapshot with $m=+1$, $J_2=0$ and $T=0.02$.
  }
 \label{fig:noisy_reconstructions}
\end{center}
\end{figure}

\subsubsection{Training IV: random training data}
\label{sec:trainingIV}

To check whether learning the order of low temperature is a consequence of the finite-size effects in the data,
we train a CAE with (pseudo-)random arrays of zeros and ones, which emulate an infinite temperature in the Ising spin system.
We denote this training as training IV.
The right panel of Fig. \ref{fig:noisy} shows the R.E. as a function of temperature for spin configurations with $J_2=0$.
The CAE can reconstruct approximately both the ordered and disordered spin configurations with $J_2=0$
(maximum R.E. is roughly $0.009$). Nonetheless, the transition is still visible and the splitting is again present.

A similar splitting is observed in the latent variable of a fully connected autoencoder\cite{alexandrou2019} with a single unit layer in
the non-frustrated square lattice, or in the largest principal component $Z_1$ of PCA.

The CAE in training IV is capable of reproducing in good approximation all the spin configurations of the $J_1-J_2$ model in the honeycomb lattice,
as shown in Fig. \ref{fig:noisy_reconstructions}.

\begin{figure}[t!]
\begin{center}
  \includegraphics[width= \linewidth]{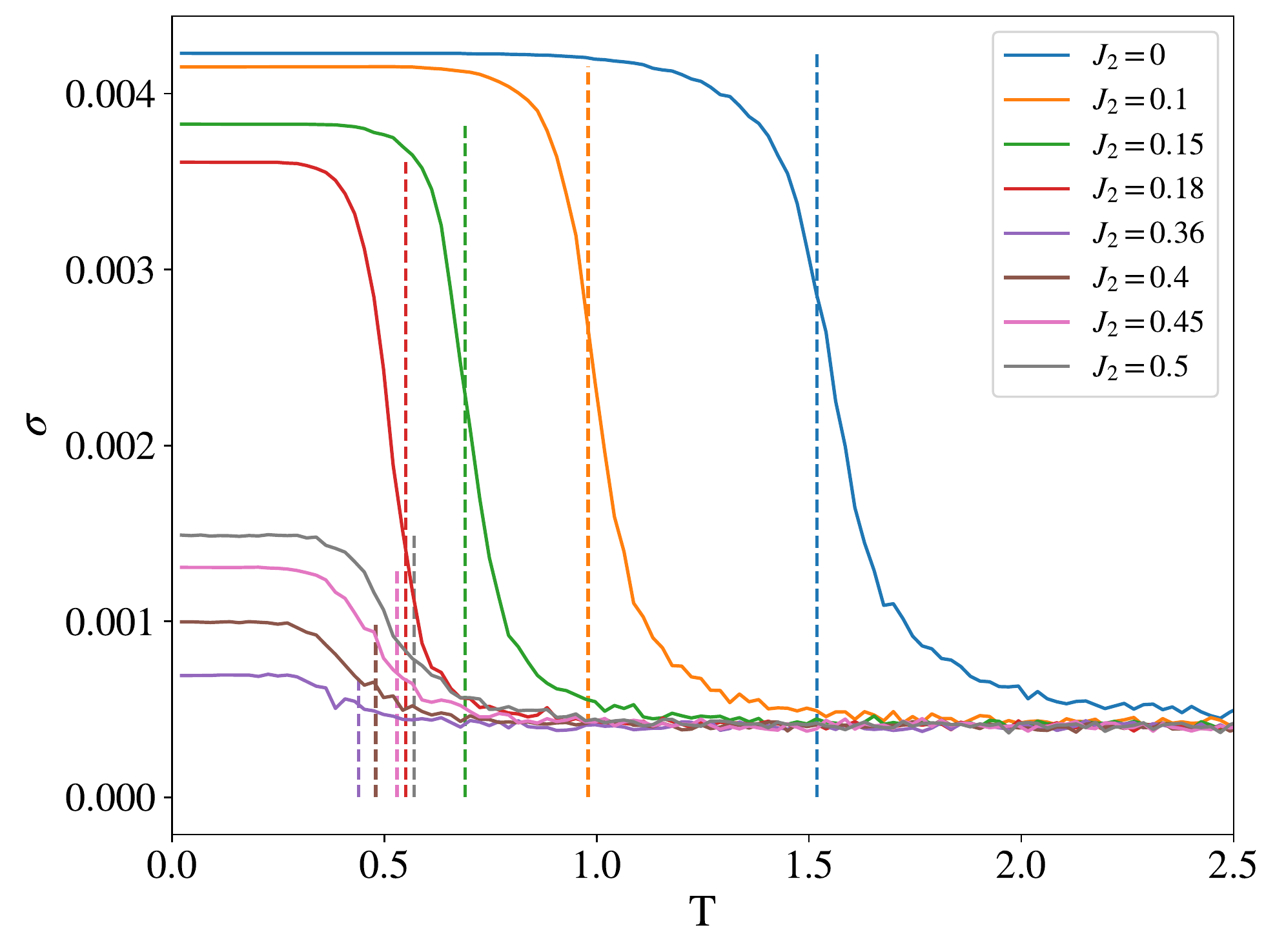}
 \caption{
 Standard deviation in R.E.  as a function of temperature for several values of $J_2$, computed with the CAE from to training IV. In dashed lines, 
 the transition temperatures corresponding to specific heat maxima in the Monte Carlo simulation. 
 }
 \label{fig:noisy_sigmas}
\end{center}
\end{figure}

Fig. \ref{fig:noisy_sigmas} shows the standard deviations in the mean R.E. as a function of temperature for several values of $J_2$ evaluating the CAE from training IV. For $J_2<1/4$, the R.E. presents bifurcations at the ordering temperatures as in Fig. \ref{fig:noisy} (not shown here), which correspond to the high variance at low temperatures in Fig. \ref{fig:noisy_sigmas}. For $J_2>1/4$ there are no mean R.E. changes larger than the standard deviation at any temperature.
The R.E. dispersion increases around the transition temperatures (dashed lines) for all $J_2$, and all R.E. dispersions converge to the same value at high temperatures.

For low temperatures, it is clear that each dispersion curve goes into a well-defined plateau with a value greater than the high temperature one.
This allows to separate high and low-temperature data far away from these inflection points and label data with this information to perform a supervised
classification as in Fig. \ref{fig:PD}.

\subsubsection{Higher ground state degeneracy }
\label{sec:Lifting-deg-random-data}

As we have observed previously, Figs. \ref{fig:colored_spliting}, and \ref{fig:noisy}  present a splitting corresponding to the two-fold degeneracy of
the ground state, due to the $Z_2$ symmetry of the Hamiltonian.
Here naturally the question arises whether the R.E. calculated by a CAE could be divided into $ n $ branches in case the ground state had a degeneracy $ n $.

\begin{figure}[t!]
\begin{center}
\includegraphics[width= .9\linewidth]{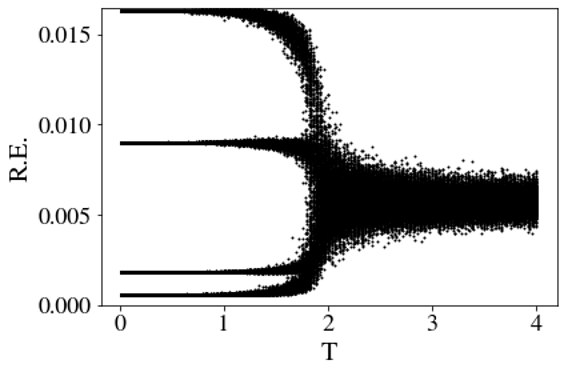}
 \caption{
 Reconstruction error vs temperature for the 400 realizations of the square lattice with $J_2/J_1=0.9$, computed with the CAE corresponding to training IV.
 }
 \label{fig:cuatrifurcation}
\end{center}
\end{figure}

For this reason, we briefly explore a system with a four-fold degeneracy. In Fig. \ref{fig:cuatrifurcation} we show the R.E. computed
with the CAE from training IV on data from the antiferromagnetic $J_1-J_2$ Ising model on the square lattice, with $J_2/J_1=0.9$.
It is well known that for  this vale,  the ground state is four-fold degenerate and exhibits a stripe-type
long-range order\cite{Andreas-Frustrated-Ising-Square-1,Andreas-Frustrated-Ising-Square-2}. \\
It is possible that in another equivalent but different training the R.E. plot shows $m$ branches with $m<n$.
For this reason, it is important to repeat the training procedure and study the different possible outcomes.
Furthermore, it is important to count the number of snapshots within each branch, constructing a histogram in R.E. for the minimum temperature value.
In this histogram (not shown here), if the ground state is $n-$degenerated and the data set is correctly balanced the bars have equal heights if $m=n$
or different heights if some of the R.E. branches cannot be discriminated.

\section{Summary and Conclusions}
\label{sec:conclusions}

In this work, we have employed unsupervised learning methods to classify phases in classical frustrated antiferromagnets, in particular the frustrated Ising model on the honeycomb lattice.

We trained convolutional autoencoders to discriminate between different magnetic phases, choosing distinct training sets in the phase space of the system. We monitor the reconstruction error of the CAE to see where the phase transitions occur.
Starting from the unfrustrated system, i.e., $J_2=0$, we can compare the R.E. curve with the analytical critical temperature of the system and see that the R.E. inflection point coincides with this temperature.
This criterion to separate two phases can be extended to the rest of the $J_2$ values and allows to construct the $T-J_2$ phase diagram of the system. Furthermore, we validate this approach using a CNN that performs a classification using labels that are obtained using the CAE, showing an interesting interplay between neural network architectures.
Additionally, we compare the transitions determined by both methods with the maximum of the specific heat obtained by means of Monte Carlo simulations, finding a very good agreement between the three methods. \\
Later on, we show how the R.E. can lift the 2-degeneracy of the N\'eel ground state, by splitting into two branches, for different training choices. We highlight that for case IV the CAE was trained to reproduce random matrices filled with zeros and ones, not generated by the Monte Carlo, although they correspond to an infinite temperature in the model.
In this case, the CAE can reproduce with low R.E. spin configurations from all the phase space of the model, and it is possible to monitor the R.E. standard deviation, instead of the R.E. mean value, to detect phase transitions.
Finally, the CAE corresponding to infinite temperature, case IV,  was used to reproduce spin configurations of the $J_1-J_2$ antiferromagnetic Ising model on the square lattice with $J_2/J_1=0.9$, to show that in this case, the R.E. can split in four, which is the ground state degeneracy in that system.

As possible perspectives we can mention firstly the anomaly detection using CAEs in frustrated quantum magnets, searching for an input with the lowest computational cost that allows discriminate in an unsupervised fashion complexly correlated phases such as quantum spin liquids.
Secondly, non-linear low dimensional representations of many-body quantum systems\cite{hubbard-unsupervised2018} are a very interesting topic since exponentially large Hilbert spaces could be approximate by smooth manifolds with a much lower number of variables.

\section*{Acknowledgments}
This research was partially supported by CONICET (P-UE 22920170100066CO), UNLP (Grant No. 11/X678, 11/X806). 
We thank I. Corte for allow us to use her data set on the square lattice. The neural network calculations were performed with tensorflow.

\section{Appendix}

\subsection*{Code availability}
The data sets and codes of this work are available from the corresponding author upon request.
\subsection{Monte Carlo simulation}
\label{sec:MC}
The Monte Carlo generation of data, using a Metropolis Algorithm and
single-spin-flip dynamics\cite{newman1999monte}, is performed as follows. For each value of $J_2$, we make 400 independent simulations starting from the high-temperature phase (initial temperature: $T_0=4.5$). A set of 200 evenly-spaced temperature values is obtained from the range $[0.02,T_0]$. For each temperature, the spin configuration and the temperature are saved once equilibrium is reached. Thus, our dataset for each value of $J_2$ consists of 80000 samples or images.
The size of the system is 900 sites and the simulation was performed using periodic boundary conditions.

\subsection{Principal Component Analysis}
\label{sec:PCA}

The PCA \cite{wang2016discovering,wang2017machine}consists in performing a reduction of the dimensionality of given a dataset, while preserving as much information as possible, by finding
new variables called the principal components. These principal components are linear functions of those in the original dataset, that successively maximize variance and that are uncorrelated with each other.
Finding these variables reduces to find the eigenvalues and eigenvectors of the sample covariance matrix associated with the dataset. Dimensional reduction is achieved by keeping only the components with the largest eigenvalues of the sample covariance matrix.
The off-diagonal elements of the sample covariance matrix are the covariances between spins, which are related to the correlation between them.\\
Non-linear representations can be obtained by using Kernel PCA\cite{wang2018machine}

\begin{figure}[t!]
\begin{center}
  \includegraphics[width= \linewidth]{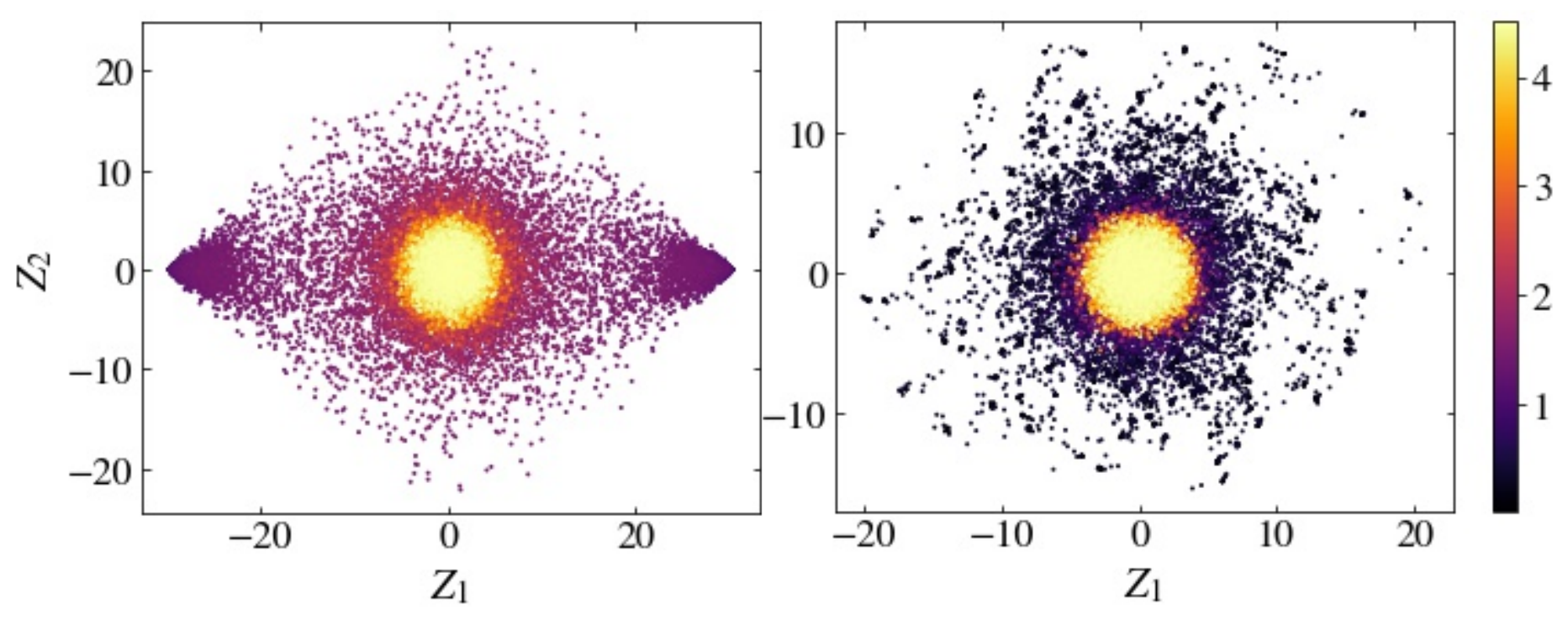}
 \caption{Two dimensional PCA representation of the system with $J_2=0$ on the left, and $J_2=1/2$ on the right.
 Each point corresponds to a spin configuration, and the color scale corresponds to the  temperature.}
 \label{fig:PCA_latent}
\end{center}
\end{figure}

\begin{figure}[t!]
\begin{center}
  \includegraphics[width= \linewidth]{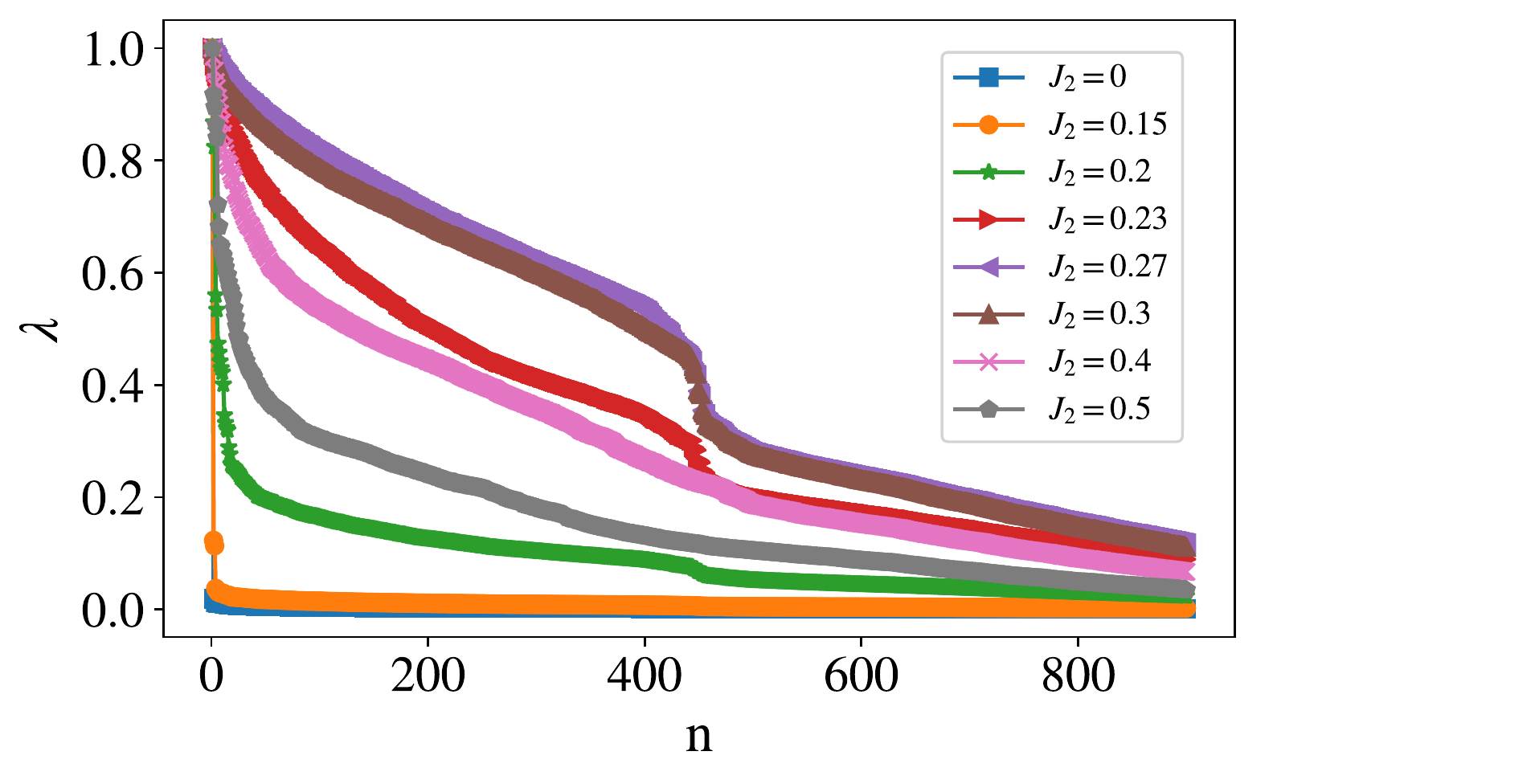}
 \caption{Principal components $\lambda$ of data, for several values of $J_2$.
 For each value of $J_2$ the 900 principal components are normalized to the largest one, and $1\leq n \leq 900$ orders them in decreasing magnitude.}
 \label{fig:PCA_evals}
\end{center}
\end{figure}

In Fig. \ref{fig:PCA_latent} we show two-dimensional representations computed with PCA	 for the Ising model with $J_2=0$ and $J_2=1/2$. 
Each spin configuration $x_n$, $1\leq n \leq 80000$, is vectorized, i.e., inserted in a 1D column vector of $N=900$ components which is used as input for the PCA.
It can be seen that high-temperature spin configurations are mapped to a disk
centered at the origin as they are uncorrelated. Nonetheless, we see that at low temperatures linear PCA fails to cluster low-temperature data in the frustrated model, whereas N\'eel ordered configurations converge to two well-defined points in these coordinates,
with $Z_1\approx \pm 30$ and $Z_2=0$. Note that in the latter case a single variable, $Z_1$, represents the ordered phase, that has only two types of ordered configurations, due to the $Z_2$ symmetry of the Hamiltonian.

Fig.  \ref{fig:PCA_evals} shows the 900 PCA eigenvalues for different values of frustration.
Each eigenvalue list is normalized to its maximum value for each $J_2$.
For $J_2=0$ (or $J_2 \ll 1/4$) there is only one relevant principal component.
This component corresponds to the order parameter of the N\'eel phase, which is a linear combination of the spins.
It can be seen that frustration increases the number of relevant principal components. It is interesting to remark that the area under the curve is maximum for $J_2$ values around $J_2=1/4$, where the system has its low-temperature transition from the N\'eel phase to another phase without long-range order\cite{Ising-Honeycomb-frustrated}.
The tendency for a larger number of principal components to be relevant around $J_2=1/4$ takes place where the system is maximally frustrated.

\subsection{Mapping the honeycomb lattice to a square array}
\label{sec:honey_map}
\begin{figure}[H]
\begin{center}
 \includegraphics[width=.6\linewidth]{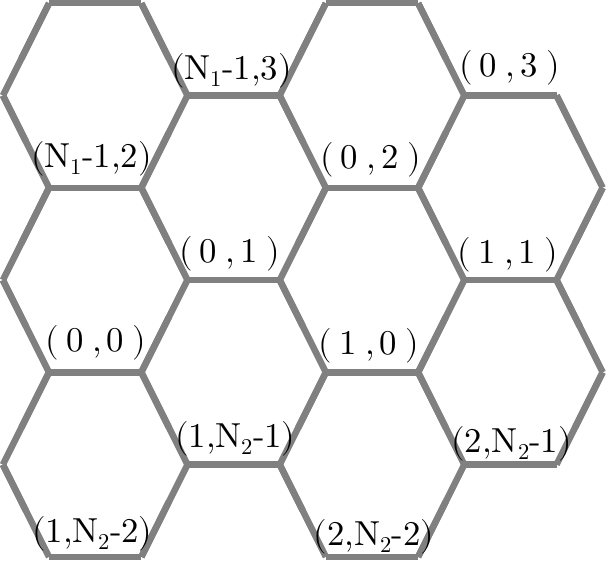}
\end{center}
\caption{Honeycomb lattice. Ordered pairs correspond to the indexation for each cell used in this work. }
\label{fig:honeycomb}
\end{figure}

 Each unit cell in the honeycomb lattice is indexed by two 
 integers $(i,j)$, where $0\leq i<N_1=30$ and $0 \leq j<N_2=15$ as we show in Fig. \ref{fig:honeycomb}.

We map the honeycomb lattice to a square $30 \times 30 $ array $A$  as follows:
each spin in the honeycomb lattice is indexed by three integers in the 3-rank tensor $S_{ij}^k$, where $i$ and $j$ determine the unit cell $(i,j)$, and $k=0$ or $k=1$ corresponds to the left or right spin in the unit cell, respectively.
Then, we can construct the square $30 \times 30 $ array $A$ by taking

\begin{equation}
 A_{mn}=S_{m \floor*{n/2}}^{mod(n,2)}
 \label{eq:map}
\end{equation}
with $0\leq m,n<30$, being
$mod(n,2)$ the rest in the division of $n$ by two, and $\floor*{n/2}$ the floor function which gets the integer part in the division.

\subsection{Architectures}
\label{sec:arquitecturas}
\subsubsection{Convolutional Autoencoders}

The encoder is made by two convolutional layers with relu activation functions, strides=2, filter size=3, padding='same', kernel and bias regularizers are of type l2 and of magnitude $10^{-4}$. The learning rate is $10^{-3}$ and batch size is 256. From the training set, we take ten percent of the data as validation data. The validation data is not used to adjust the network parameters. It is used to monitor if the network is overfitting. Each convolutional layer is followed by a Dropout layer with a dropout rate of 0.2.
The initial size of the input images is $30 \times 30$, and the feature maps generated by the encoder are of size $8 \times 8 $.

The decoder is made first by two convolutional transpose layers with the same characteristics as the encoder convolutional layers. Then, the final layer is a convolutional layer with filter size 3, a sigmoid activation function, and a single filter, that combines all feature maps in the final output.

We denote $N_1$ and $N_2$ the number of filters in the first and second layers of both the encoder and decoder, respectively. In trainings I and II  $N_1=16$ and $N_2=8$, whereas in trainings III and IV $N_1=32$ and $N_2=16$. Here we emphasize that the CAE do not have a latent space of dimension $N_2$ because the enconder produces a set of $N_2$ feature maps each one of dimension $8 \times 8 $.

In trainings I, II and III the number of epochs used is between 100 and 200. 
Training IV consisted of 2 epochs over 250000 pseudo-random snapshots.

Input data of the neural network are normalized. Up spins are ones and down spins are zeros. Then, when computing the mean square error between the CAE's output and input, the maximum possible mean square error is 1.

\subsubsection{Convolutional classifiers}

The CNN is made by two convolutional layers with relu activation functions, filter size=3, no padding, and no regularizers. The learning rate is $10^{-4}$ and the batch size is 256. Each convolutional layer is followed by a max-pooling layer with a pool-size of 2.
Next, there is a flattening layer, connected to a dense layer of 16 neurons with relu activation functions. Finally, the output layer has 2 neurons with a softmax activation function.

As the maximum temperature in our Monte Carlo data is $4.53$ and some transitions temperatures are as low as $ T \approx 0.3$, the training sets are very unbalanced. To balance our datasets in the classification we drop some high temperatures configurations.
For $J_2>1/4$ the training sets contain data with $0.02<T<1$, and for $J_2<1/4$ the training sets contain data with $0.02<T<3$. From each training set, we take ten percent of the data as validation data. In every training, validation accuracy is higher than 0.99.

\subsection{Inflection point estimation}
\label{sec:SG}

To obtain an estimation of the inflection point in the R.E. results we first compute the mean value over the 400 realizations of the system. 
Then, we apply a Savitzky-Golay filter from the scipy library, which is used to smooth the data, by fitting several local low degree polynomials.
In this work, we use a window length of 21 and a polynomial order of 3. Finally, we take the numerical derivative and we find its maximum.


\bibliography{refs-2020}


\end{document}